\documentclass[twocolumn,showpacs,superscriptaddress,amsmath,amssymb,nofootinbib]{revtex4-1}
\usepackage{graphicx}
\usepackage{epsfig}
\usepackage{dcolumn}
\usepackage{bm}
\usepackage{overpic}
\usepackage{color}
\usepackage{subfigure}
\usepackage{threeparttable}
\usepackage{lineno}
\usepackage[colorlinks,linkcolor=blue,anchorcolor=blue,citecolor=blue]{hyperref}

\def \lmdc {\Lambda_{c}}

\def \lambdacp   {\Lambda_{c}^{+}}
\def \lmdcp {\Lambda_{c}^{+}}

\def \lambdacm   {\bar{\Lambda}_{c}^{-}}
\def \lmdcm {\bar{\Lambda}_{c}^{-}}

\def \lmdetapi {\Lambda\eta\pi^{+}}

\def \sigstareta {\Sigma^{*+}\eta}

\def \sigstar {\Sigma^{*+}}

\def \ks    {K_{S}^{0}}
\def \pkpi  {pK^{-}\pi^{+}}

\def \ee   {e^+e^-}
\def \gev  {\mbox{GeV}}

\def \gevcc{\mbox{GeV/$c^2$}}
\def \mev  {\mbox{MeV}}
\def \mevcc{\mbox{MeV/$c^2$}}
\def \ipb  {\mbox{pb$^{-1}$}}

\def \vz   {V_{z}}
\def \BR   {\mathcal{B}}

\def \L {\mathcal{L}}

\def \epem {e^+e^-}
\def \ee{e^+e^-}

\def \piz  {\pi^0}
\def \pip  {\pi^+}
\def \pim  {\pi^-}

\def \eff  {\varepsilon}

\def \dedx {\mbox{d}E/\mbox{d}x}
\def \mbc {M_{\rm{BC}}}
\def \deltaE {\Delta E}

\begin{document}
\title{\bf\boldmath Measurement of the absolute branching fractions of $\lmdcp \to \lmdetapi$ and $\Sigma(1385)^{+}\eta$}
\author{\small
M.~Ablikim$^{1}$, M.~N.~Achasov$^{10,d}$, S. ~Ahmed$^{15}$, M.~Albrecht$^{4}$, M.~Alekseev$^{55A,55C}$, A.~Amoroso$^{55A,55C}$, F.~F.~An$^{1}$, Q.~An$^{52,42}$, Y.~Bai$^{41}$, O.~Bakina$^{27}$
% ...
\vspace{0.2cm}
(BESIII Collaboration)\\
\vspace{0.2cm} {\it
$^{1}$ Institute of High Energy Physics, Beijing 100049, People's Republic of China\\
% ...
$^{4}$ Bochum Ruhr-University, D-44780 Bochum, Germany\\
% ...
$^{10}$ G.I. Budker Institute of Nuclear Physics SB RAS (BINP), Novosibirsk 630090, Russia\\
% ...
$^{15}$ Helmholtz Institute Mainz, Johann-Joachim-Becher-Weg 45, D-55099 Mainz, Germany\\
% ...
$^{27}$ Joint Institute for Nuclear Research, 141980 Dubna, Moscow region, Russia\\
% ...
$^{41}$ Southeast University, Nanjing 211100, People's Republic of China\\
$^{42}$ State Key Laboratory of Particle Detection and Electronics, Beijing 100049, Hefei 230026, People's Republic of China\\
% ...
$^{52}$ University of Science and Technology of China, Hefei 230026, People's Republic of China\\
% ...
$^{55}$ (A)University of Turin, I-10125, Turin, Italy; (B)University of Eastern Piedmont, I-15121, Alessandria, Italy; (C)INFN, I-10125, Turin, Italy\\
% ...
$^{d}$ Also at the Novosibirsk State University, Novosibirsk, 630090, Russia\\
% ...
}\vspace{0.4cm}}

\date{\today}
\begin{abstract}
	We study the decays $\lmdcp \to \lmdetapi$ and $\Sigma(1385)^{+}\eta$ based on $\lmdcp\lmdcm$ pairs produced in $\ee$ collisions at a center-of-mass energy of $\sqrt{s}=4.6~\gev$, corresponding to an integrated luminosity of $567\;\ipb$. The data sample was accumulated with the BESIII detector at the BEPCII collider.
	The branching fractions are measured to be $\BR(\lmdcp \to \lmdetapi)=(1.84\pm0.21({\rm{stat.}})\pm0.15({\rm{syst.}}))\%$ and $\BR(\lmdcp \to \Sigma(1385)^{+}\eta)=(0.91\pm0.18({\rm{stat.}})\pm0.09({\rm{syst.}}))\%$, constituting the most precise measurements to date.
\end{abstract}
\maketitle

	
\section{\boldmath Introduction}
	Since the charmed baryon ground state $\lmdcp$ was first observed at the Mark II experiment in 1979~\cite{1980GSAbrams}, progress in the studies of charmed baryon decays was relatively slow both theoretically and experimentally due to the limits of the factorization approach in complicated three quark systems~\cite{1992HYCheng} and the lack of experimental data, respectively. Therefore, more efforts in studying hadronic decays of the $\lmdcp$ are useful to understand the internal dynamics of charmed baryons.
	
	Theoretically, in Ref.~\cite{2016JJXie}, the decay $\lmdcp\to\lmdetapi$ was pointed out as an ideal process to study the $a_0(980)$ and $\Lambda(1670)$, because the final states $\eta\pip$ and $\Lambda\eta$ are in pure isospin $I=1$ and $I=0$ combinations. Also in Ref.~\cite{2015Kenta}, resonances $\Lambda(1405)$ and $\Lambda(1670)$ have been studied in $\Lambda\eta$ combinations, and in Ref.~\cite{2017JJXie}, several $\Sigma^{*}$ states including possible pentaquark state $\Sigma_{1/2^-}(1380)$ and resonance $\Sigma(1385)$  have been studied in $\Lambda\pip$ combinations.
	Experimentally, the decays $\lmdcp\to\lmdetapi$ and $\Sigma(1385)^{+}\eta$~\footnote{For simplicity, we use the symbol $\sigstar$ to represent $\Sigma(1385)^+$ resonance throughout this paper.} have been studied at the CLEO experiment in 1995~\cite{1995RAmmar} and 2003~\cite{2003DCronin}. The branching fractions (BFs) for both channels are measured relative to $\BR(\lmdcp\to\pkpi)$. After scaling with the average $\BR(\lmdcp\to\pkpi)$ given by the Particle Data Group (PDG)~\cite{2016PDG}, the absolute BFs are estimated as $\BR(\lmdcp\to\lmdetapi)=(2.2\pm0.5)\%$ and $\BR(\lmdcp\to\sigstareta)=(1.22\pm0.37)\%$, with large uncertainties at the 20\% and 30\% level, respectively. 
	
	In this paper, we present an improved measurement of the absolute BFs of the $\lmdcp \to \lmdetapi$ and study the intermediate state $\sigstar$ in the three-body decay. The measurements are based on a $\lmdcp\lmdcm$ pair data sample produced in $\epem$ collisions at a center-of-mass energy $\sqrt s=4.6\;\gev$~\cite{2016Energy}, corresponding to an integrated luminosity of 567 $\ipb$~\cite{2015Lumi}. The sample was collected by the $\uchyph=0$BESIII detector~\cite{2009Detector} at the Beijing Electron Positron Collider (BEPCII)~\cite{Yu:IPAC2016-TUYA01}. The collision energy is just above the mass threshold for the production of $\lmdcp\lmdcm$ pairs, providing a very clean environment without the production of additional hadrons. Taking advantage of this and the excellent performance of the $\uchyph=0$BESIII detector, a single-tag method ($i.e.$ only one $\lmdc$ of the $\lmdcp\lmdcm$ pair is reconstructed in each event and the other $\bar{\Lambda}_c$ is assumed in the recoil side) is used in the analysis, in order to improve the detection efficiency and acquire more $\lmdcp$ candidates. The single-tag method is valid under the condition that $\lmdcp$ and $\lmdcm$ are always produced in pairs. In this paper, CP violation will be neglected which is reasonable from the studies on the current statistics-limited data set; thus the charge conjugate states are always implied unless mentioned explicitly.
	
\section{\boldmath BESIII experiment and Monte Carlo simulation}
	The $\uchyph=0$BESIII detector is a magnetic spectrometer located at the $\uchyph=0$BEPCII collider. The cylindrical core of the BESIII detector consists of a helium-based multilayer drift chamber (MDC), a plastic scintillator time-of-flight system (TOF), and a CsI(Tl) electromagnetic calorimeter (EMC), which are all enclosed in a superconducting solenoidal magnet providing a 1.0~T magnetic field. The solenoid is supported by an octagonal flux-return yoke with resistive plate counter muon identifier modules interleaved with steel. The acceptance of charged particles and photons is 93\% over $4\pi$ solid angle. The charged-particle momentum resolution at $1~{\rm GeV}/c$ is $0.5\%$, and the $\dedx$ resolution is $6\%$ for the electrons from Bhabha scattering. The EMC measures photon energies with a resolution of $2.5\%$ ($5\%$) at $1$~GeV in the barrel (end cap) region. The time resolution of the TOF barrel part is 68~ps, while that of the end cap part is 110~ps. More detailed descriptions can be found in Refs.~\cite{2009Detector,Yu:IPAC2016-TUYA01}.
	
	Simulated samples produced with the {\sc geant4}-based~\cite{geant4} Monte Carlo (MC) package which
includes the geometric description of the BESIII detector~\cite{GDMLMethod,BesGDML} and the detector response, are used to determine the detection efficiency and to estimate the backgrounds. The simulation includes the beam energy spread and initial state radiation (ISR) in the $e^+e^-$
annihilations modelled with the generator {\sc kkmc}~\cite{ref:kkmc}. The inclusive MC samples consist of the production of open charm
processes, the ISR production of vector charmonium(-like) states, and the continuum processes incorporated in {\sc kkmc}~\cite{ref:kkmc}. The known decay modes are modelled with {\sc evtgen}~\cite{ref:evtgen} using branching fractions taken from the
Particle Data Group~\cite{2016PDG}, and the remaining unknown decays from the charmonium states with {\sc lundcharm~\cite{ref:lundcharm}}. The final state radiations (FSR) from charged final state particles are incorporated with the {\sc photos} package~\cite{photos}. For the production of $\epem \to \lmdcp\lmdcm$ signal MC samples, which are used to estimate the detection efficiencies, the observed cross sections~\cite{2018CrossSection} are taken into account in simulating ISR, and the observed kinematic behavior is considered when simulating $\lmdcp$ decays.	

\section{\boldmath Event Selection}
	Charged particle tracks are reconstructed from hits in the MDC, and are required to have a polar angle $\theta$ with respect to the beam direction satisfying $|\cos\theta|<$ 0.93 and a distance of closest approach to the interaction point (IP) of less than $10\,\mbox{cm}$ along the beam axis ($\vz$) and less than $1 \,\mbox{cm}$ in the plane perpendicular to the beam axis, except for those used to reconstruct the $\Lambda\to p \pim$ decay.
	Particle identification (PID) for charged particle tracks combines the information from the flight time in the TOF and measurements of ionization energy loss ($\dedx$) to form a likelihood $\L(h)$ ($h=\pi, K, p$) for each hadron ($h$) hypothesis. Tracks will be identified as protons when this hypothesis is determined to have the largest PID likelihood ($\L(p)>\L(K)$ and $\L(p)>\L(\pi)$), while charged pions are differentiated from kaons by the likelihood requirement $\L(\pi)>\L(K)$. 

\begin{figure}[htbp]
\centering
	 \includegraphics[trim = 9mm 0mm 0.mm 0mm, width=0.32\textwidth]{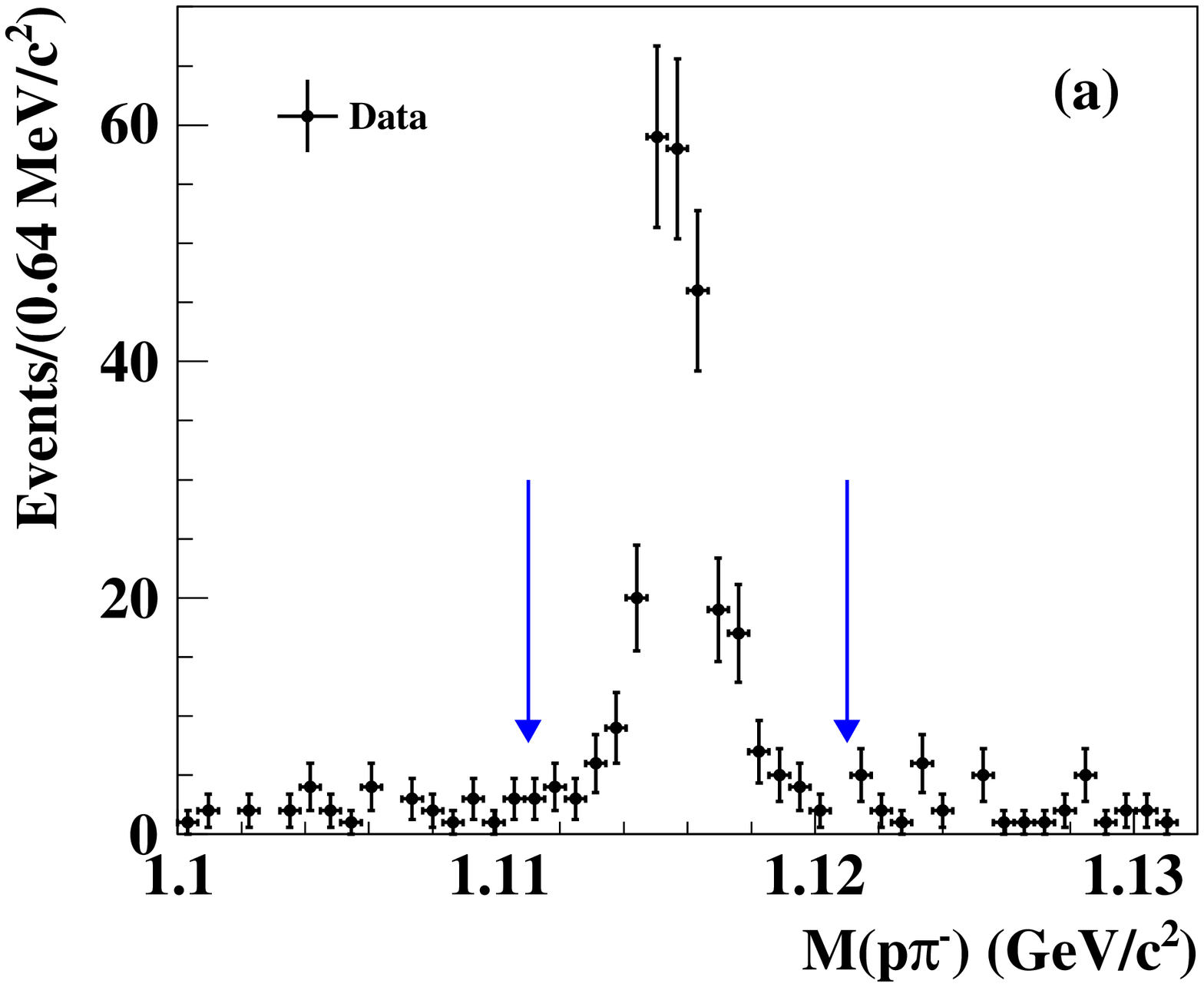}
	 \includegraphics[trim = 9mm 0mm 0.mm 0mm, width=0.32\textwidth]{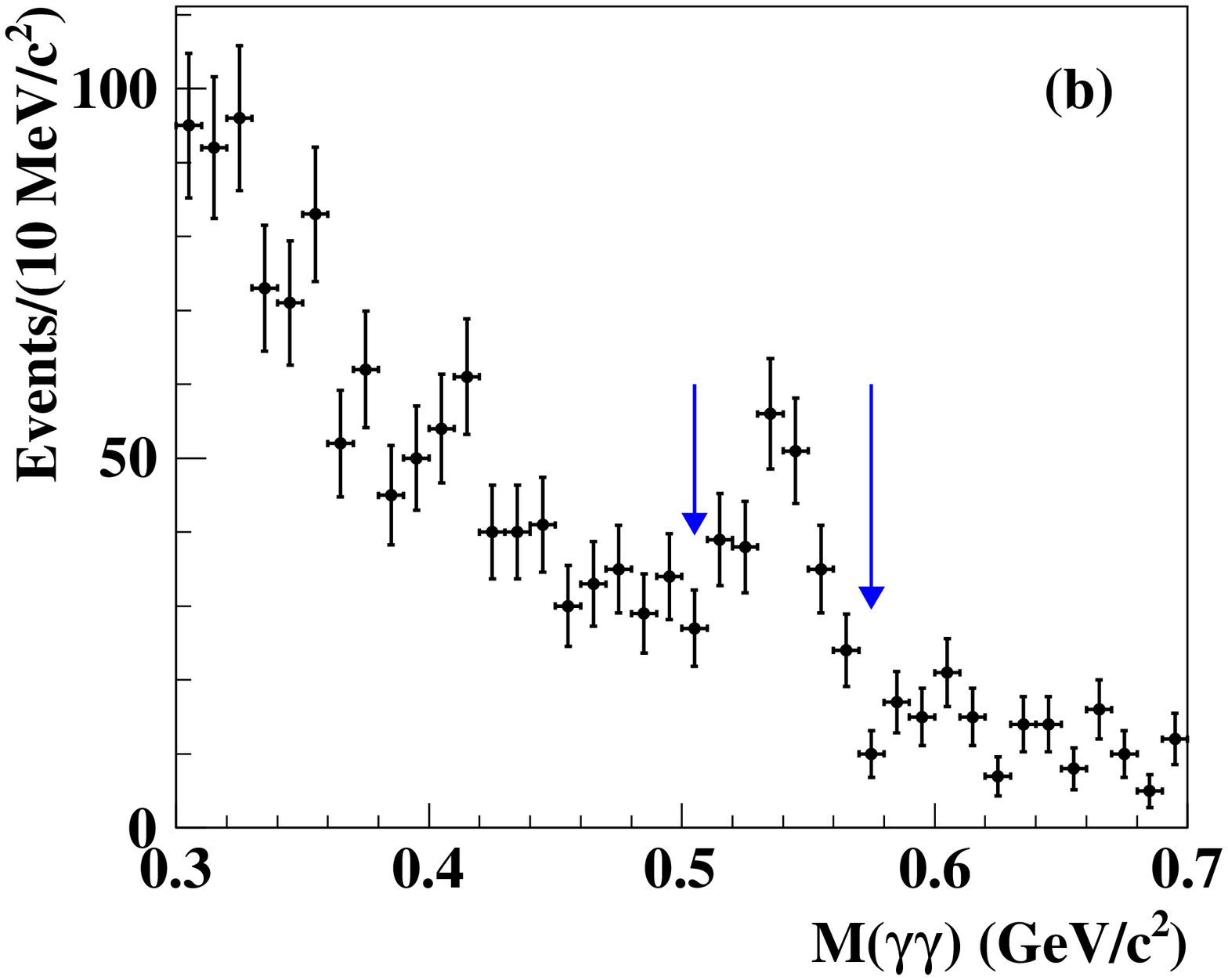}
	 \includegraphics[trim = 9mm 0mm 0.mm 0mm, width=0.32\textwidth]{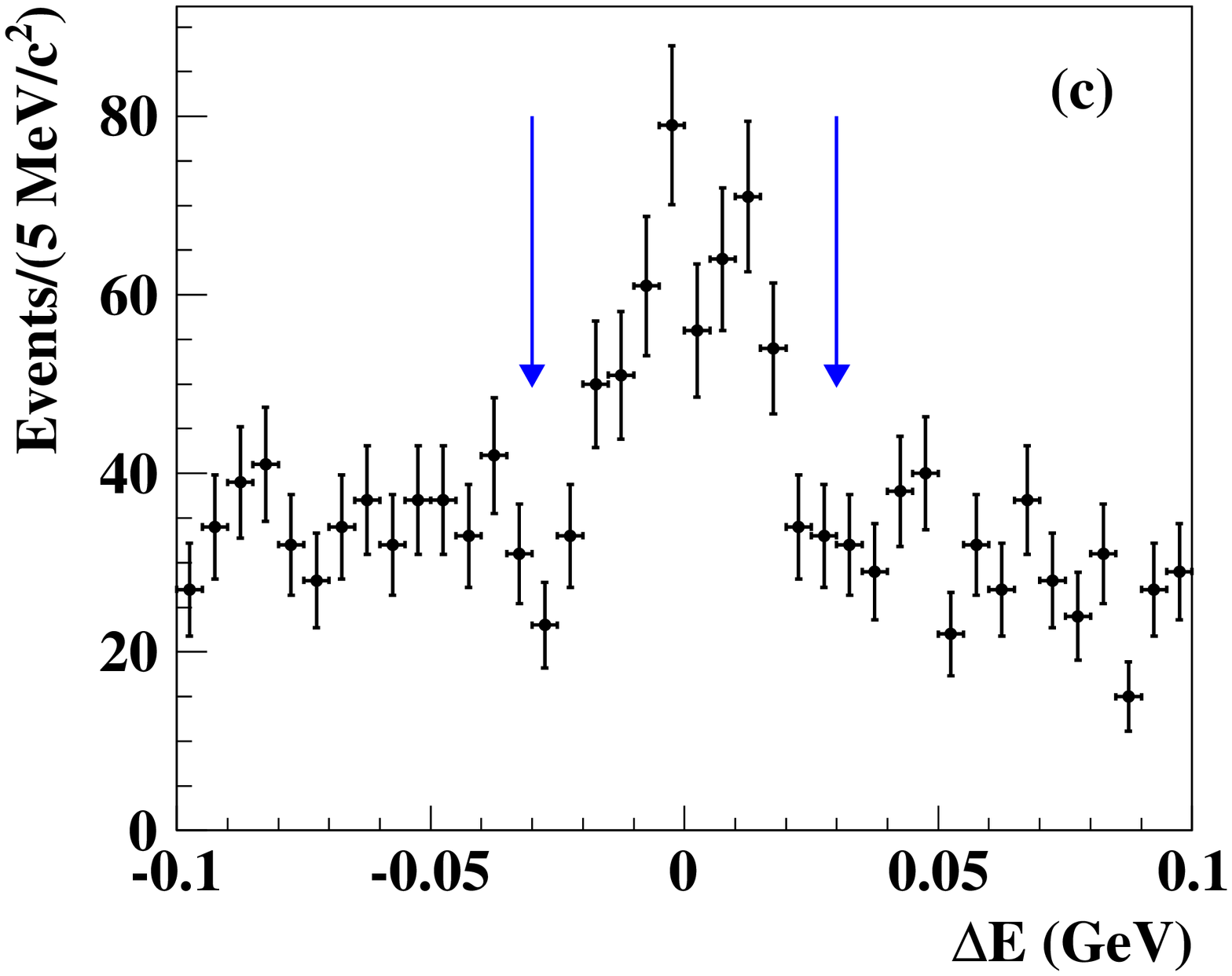}
\caption{Invariant mass spectra of the $p\pim$ pairs (a) and $\gamma\gamma$ pairs (b) used for selecting the $\Lambda$ and $\eta$ candidates, respectively, and energy difference distribution (c) for selecting the signal events candidates. The points with error bars stand for data and the arrows indicate the mass or energy difference requirement. For better illustrations of the signals in plotting, all subfigures are drawn under $\mbc$ fitting range $2.25<\mbc<2.30\;\gevcc$, while additional requirement $-0.03<\deltaE<0.03\;\gev$ are applied in subfigures (a) and (b).}
\label{fig:lmd_eta_deltaE}
\end{figure}

	Clusters with no association to a charged particle track in the EMC crystals are identified as photon candidates when satisfying the following requirements: The deposited energy is required to be larger than 25 $\mev$ in the barrel region ($|\cos\theta|<$ 0.80) or 50 $\mev$ in the end-cap region (0.86 $<|\cos\theta|<$ 0.92). To suppress background from electronic noise and showers unrelated to the events, the measured EMC time is required to be within 0 and 700 ns of the event start time. Additionally, in order to eliminate showers related to charged particle tracks, showers are required to be separated by more than 10$^\circ$ from charged particle tracks. The $\eta$ meson candidates are reconstructed from photon pairs using an invariant mass requirement of $505<M(\gamma\gamma)<575\;\mevcc$. The invariant mass spectrum of $\gamma\gamma$ pairs in data is shown in Fig.~\ref{fig:lmd_eta_deltaE}. To improve the momentum resolution, a kinematic fit constraining the invariant mass to the $\eta$ nominal mass~\cite{2016PDG} is applied to the photon pairs and the resultant energy and fitted momentum of the $\eta$ meson are used for further analysis.

Candidate $\Lambda$ baryons are reconstructed by combining two oppositely charged tracks for any pairs of $p\pim$. Those tracks are required to satisfy the polar angle requirement $|\cos\theta|<0.93$ and $\vz<20\;\mbox{cm}$ for the distance of closest approach to the IP along the beam axis. No distance constraint is applied in the plane perpendicular to the beam axis. Proton PID is required to improve the signal purity while no PID requirement is applied to the charged pion candidates. The $p$ and $\pim$ tracks are constrained to originate from a common decay vertex by requiring the $\chi^2$ of a vertex fit to be less than 100. 
Furthermore, the reconstructed momentum of the $\Lambda$ candidate is constrained to be aligned with the line joining the IP and the decay vertex, and the resultant flight distance is required to be larger than twice the fitted resolution.
A clear $\Lambda$ peak appears in the invariant mass spectrum of $p\pim$ in data, as shown in Fig.~\ref{fig:lmd_eta_deltaE}. The $p\pim$ pairs satisfying the mass requirement $1.111<M(p\pim)<1.121\;\gevcc$ are chosen as the final $\Lambda$ candidates. This requirement is chosen corresponding to $\pm\,3$ standard deviations of the reconstruction resolution around the $\Lambda$ nominal mass~\cite{2016PDG}.

	The $\lambdacp$ baryon candidates are reconstructed using all combinations of the selected $\Lambda$, $\eta$ and $\pip$ candidates. To differentiate $\lambdacp$ from background, two kinematic variables calculated in the center-of-mass system, the beam constrained mass $\mbc\equiv \sqrt{E_{\rm{beam}}^2/c^4-|\overrightarrow{p}_{\lambdacp}|^2/c^2}$, and the energy difference $\deltaE \equiv E_{\lambdacp} - E_{\rm{beam}}$ are used, where $E_{\lambdacp}$ and $\overrightarrow{p}_{\lambdacp}$ are the energy and momentum of the reconstructed $\lmdcp$ candidate respectively, and $E_{\rm{beam}}$ is the average value of the electron and positron beam energies. For a well reconstructed $\lambdacp$ candidate, $\mbc$ and $\deltaE$ are expected to be consistent with the $\lambdacp$ nominal mass and zero, respectively. Candidates are rejected when they fail the requirement of $-0.03<\deltaE<0.03\;\gev$, which corresponds to $\pm\,3$ standard deviations of the signal $\deltaE$ distribution. The $\deltaE$ distribution in data is shown in Fig.~\ref{fig:lmd_eta_deltaE}. If more than one candidate satisfies the above requirements, we select the one with the minimal $|\deltaE|$.

\section{\boldmath Signal Yield and branching fraction}
\begin{figure}[htbp]
\centering
	\includegraphics[trim = 10mm 0mm 0mm 0mm,width=0.4\textwidth]{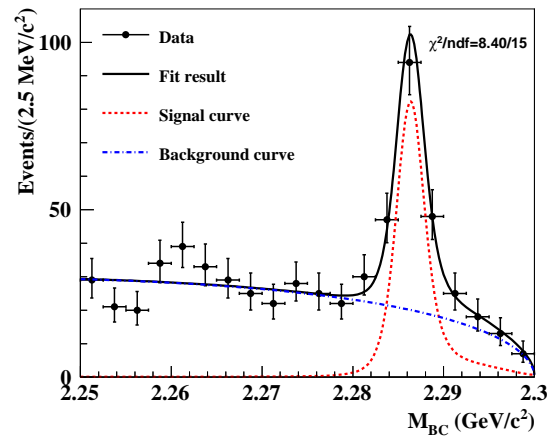}
\caption{Fit to the $\mbc$ distribution for the $\lmdcp\to\lmdetapi$ decay. The dots with error bars are data, the (black) solid curve is the fit function which is the sum of the signal shape (red dashed curve) and the background shape 
(blue dash-dotted curve). A test of goodness-of-fit with $\chi^2$ divided by the degrees of freedom is shown.}
\label{fig:lmdetapi_yield}
\end{figure}

To extract the signal yield for the $\lmdcp\to\lmdetapi$ decay, an unbinned extended maximum likelihood fit is performed to the $\mbc$ distribution in data with fitting range $2.25<\mbc<2.30\;\gevcc$, as illustrated in Fig.~\ref{fig:lmdetapi_yield}. In the fit, the signal shape is derived from the kernel-estimated non-parametric shape~\cite{keyspdf} based on signal MC samples convolved with a Gaussian function to account for the difference between data and the MC simulation caused by imperfect modeling of the detector resolution and beam-energy spread. The high mass tail in that signal shape reflects ISR effects. The parameters of the Gaussian function are free in the fit. The background shape is modeled with an $\uchyph=0$ARGUS function~\cite{1990argus} with fixed end-point $E_{\rm{beam}}$. The obtained signal yield and the corresponding detection efficiency are listed in Table~\ref{tab:sum}. The validity of the $\uchyph=0$ARGUS function to describe the background shape in the $\mbc$ spectrum is checked using the inclusive MC samples. No obvious peaking background from the decay $\lmdcp\to p \ks\eta$ with $\ks\to\pip\pim$ is observed and the influence of cross feed is neglected. The BF is calculated using 
\begin{eqnarray}
\BR(\lambdacp\to \lmdetapi) = \frac{N_{\rm{sig}}}{2\cdot N_{\lambdacp\lambdacm}\cdot\eff\cdot \BR_{\rm{inter}}},
\label{eq:BF}
\end{eqnarray}
where $N_{\rm{sig}}$ is the signal yield obtained from the $\mbc$ fit, $N_{\lambdacp\lambdacm}=(105.9\pm4.8(\rm{stat.})\pm0.5(\rm{syst.}))\times10^{3}$ is the number of $\lmdcp\lmdcm$ pairs in the data sample~\cite{2015MAblikimLambdac}, $\eff$ is the detection 
efficiency estimated using the signal MC simulation sample, and $\BR_{\rm{inter}}=\BR(\Lambda\to p\pim) \cdot \BR(\eta\to\gamma\gamma)$ is taken from the PDG~\cite{2016PDG}. The factor of 2 in the denominator takes into account the charge conjugate decay mode of the $\lmdcp$ baryon. The resultant BF and corresponding statistical uncertainty are listed in Table~\ref{tab:sum}.

  \begin{table}[htbp]
  \begin{center}
  \footnotesize
  \caption{Summary of the signal $\righthyphenmin=2$yields, the detection efficiencies, and the BFs for the different $\lambdacp$ decay modes. In the BFs, the first uncertainties are statistical, and the second are systematic.}
  \begin{tabular}{l c c c c c}
  \hline\hline
                       &  $\lmdetapi$   &  $\sigstareta$      \\ \hline
  $N_{\rm{sig}}$       &   $154 \pm 17$             &    $54 \pm 11$    \\
   $\varepsilon(\%)$    &   $15.73 \pm 0.01$           &    $12.84 \pm 0.01$           \\
  $\BR(\%)$&   $1.84 \pm 0.21\pm0.15$  &  $0.91 \pm 0.18\pm0.09$\\
  \hline\hline
  \end{tabular}
  \label{tab:sum}
  \end{center}
   \end{table}

\begin{figure*}[htbp]
\begin{center}
    \includegraphics[trim = 9mm 0mm 0.mm 0mm, width=0.32\textwidth]{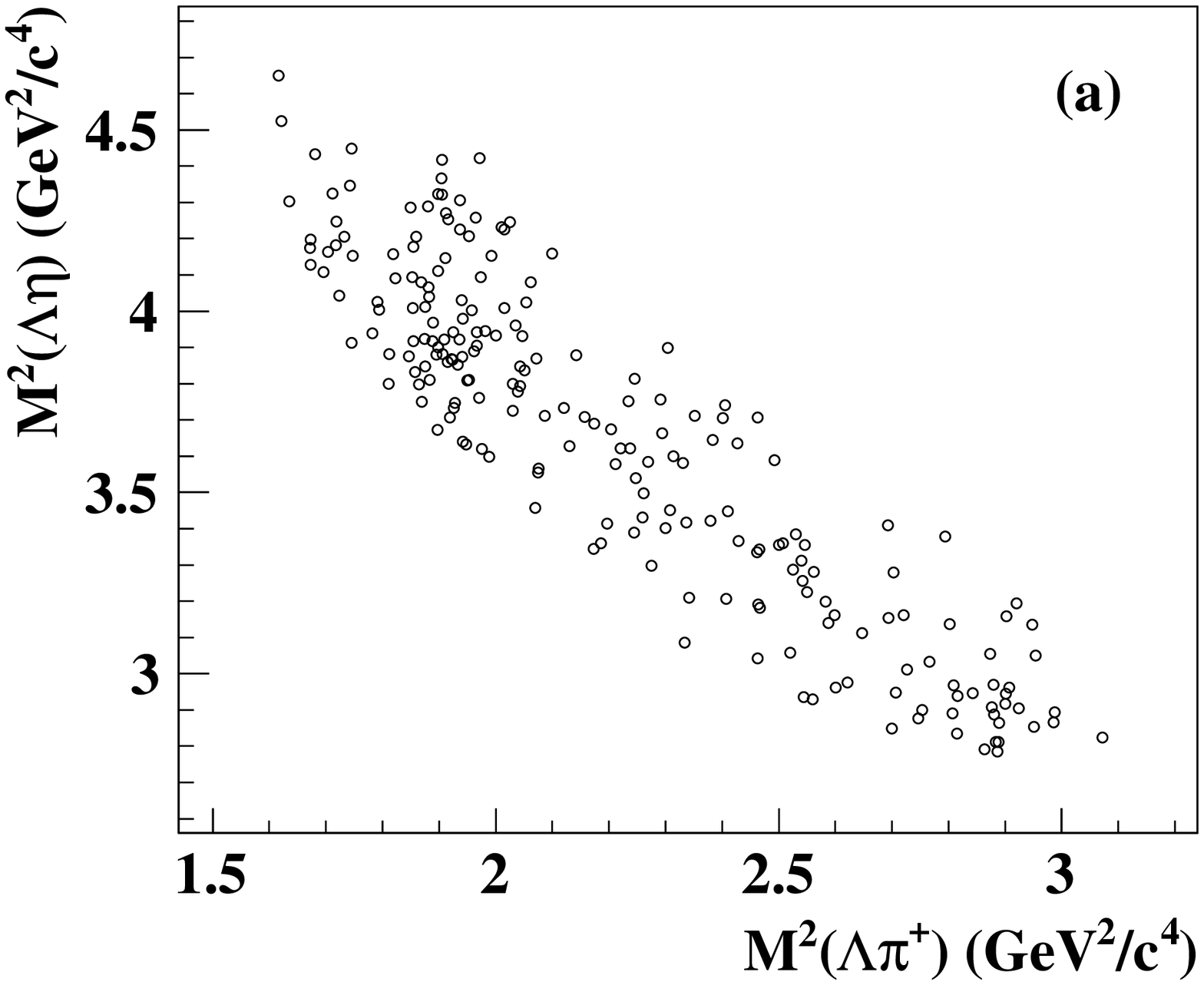} $\;\;\;\;$
    \includegraphics[trim = 9mm 0mm 0.mm 0mm, width=0.32\textwidth]{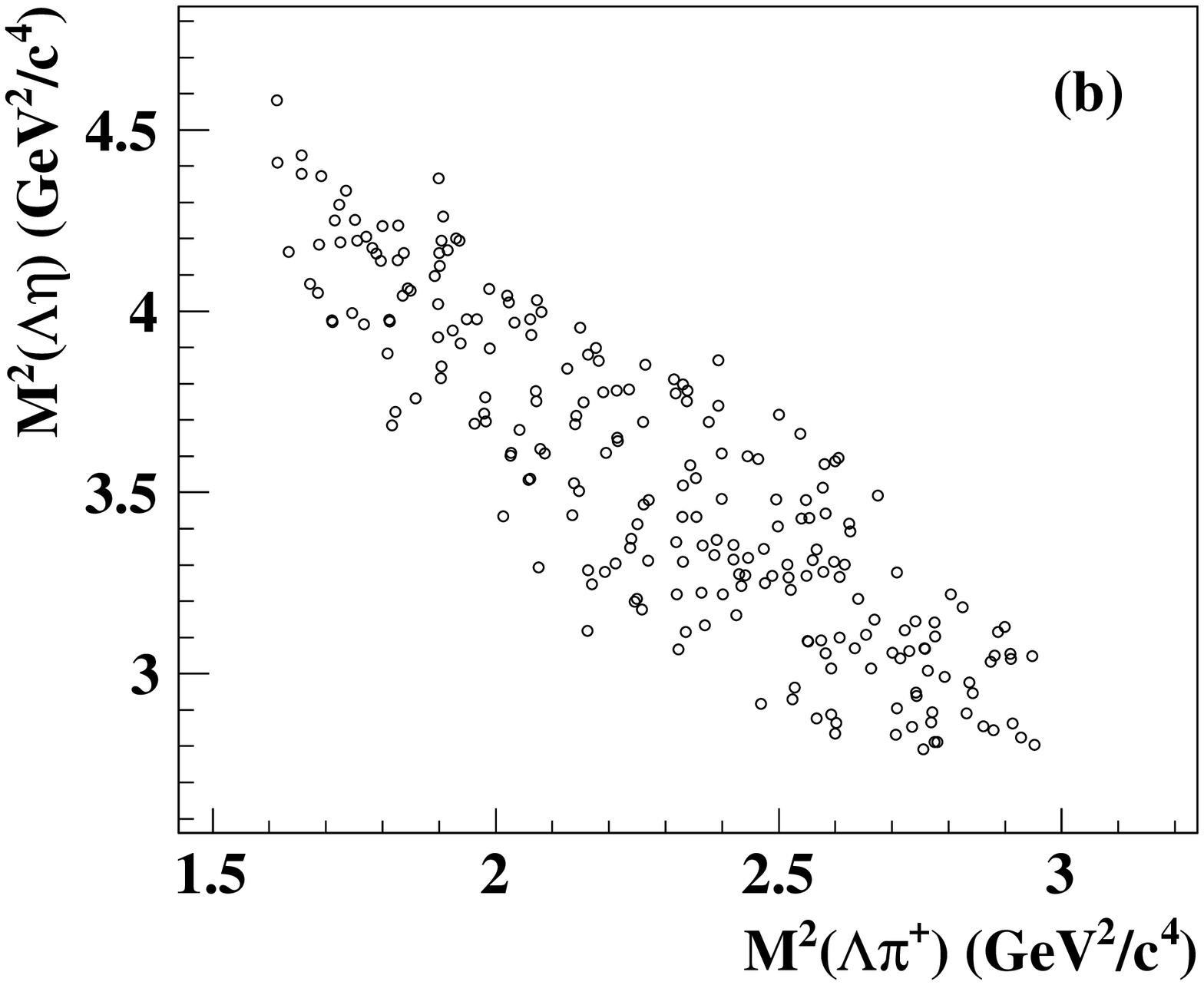}\\
    \includegraphics[trim = 9mm 0mm 0.mm 0mm, width=0.32\textwidth]{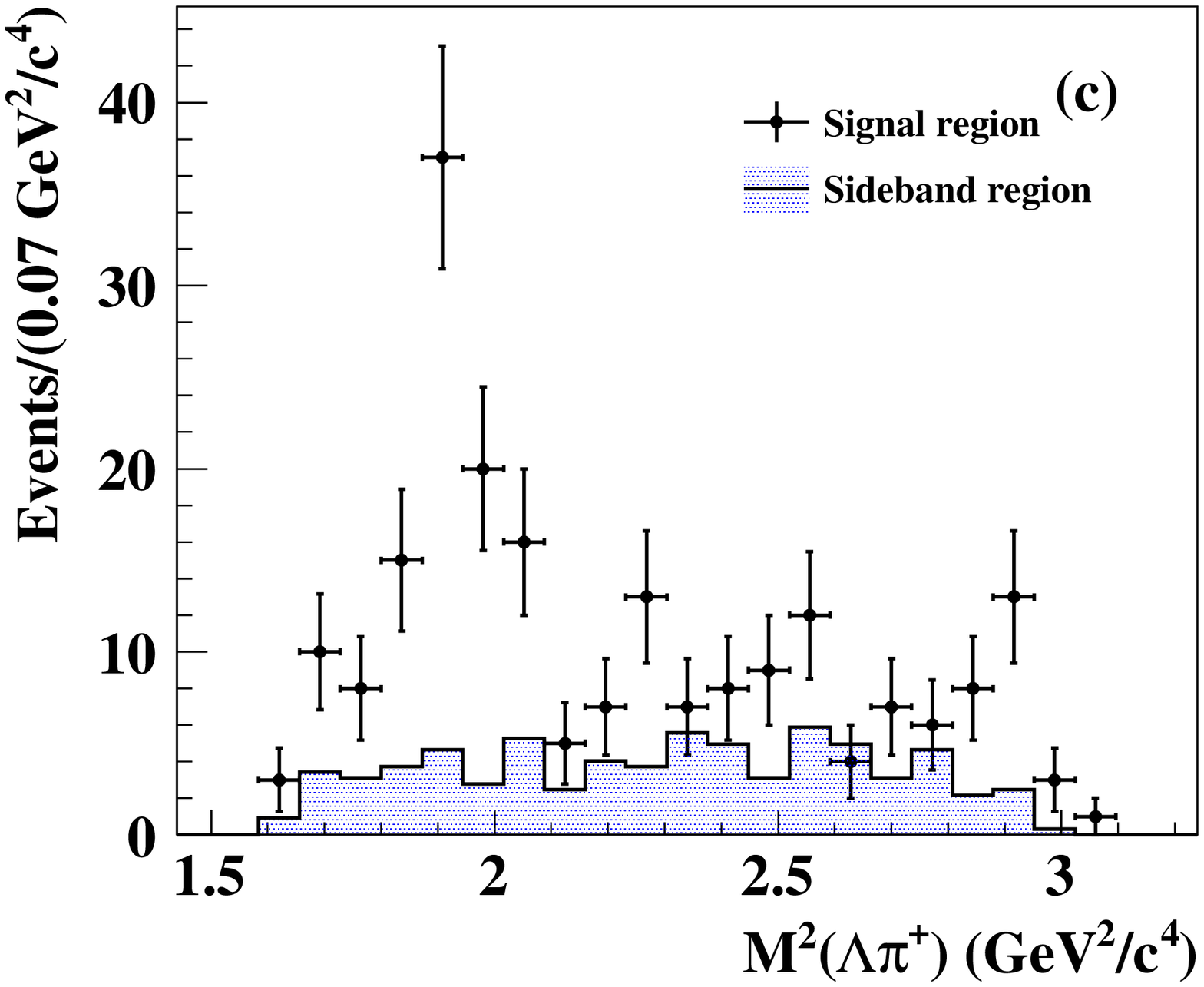}
    \includegraphics[trim = 9mm 0mm 0.mm 0mm, width=0.32\textwidth]{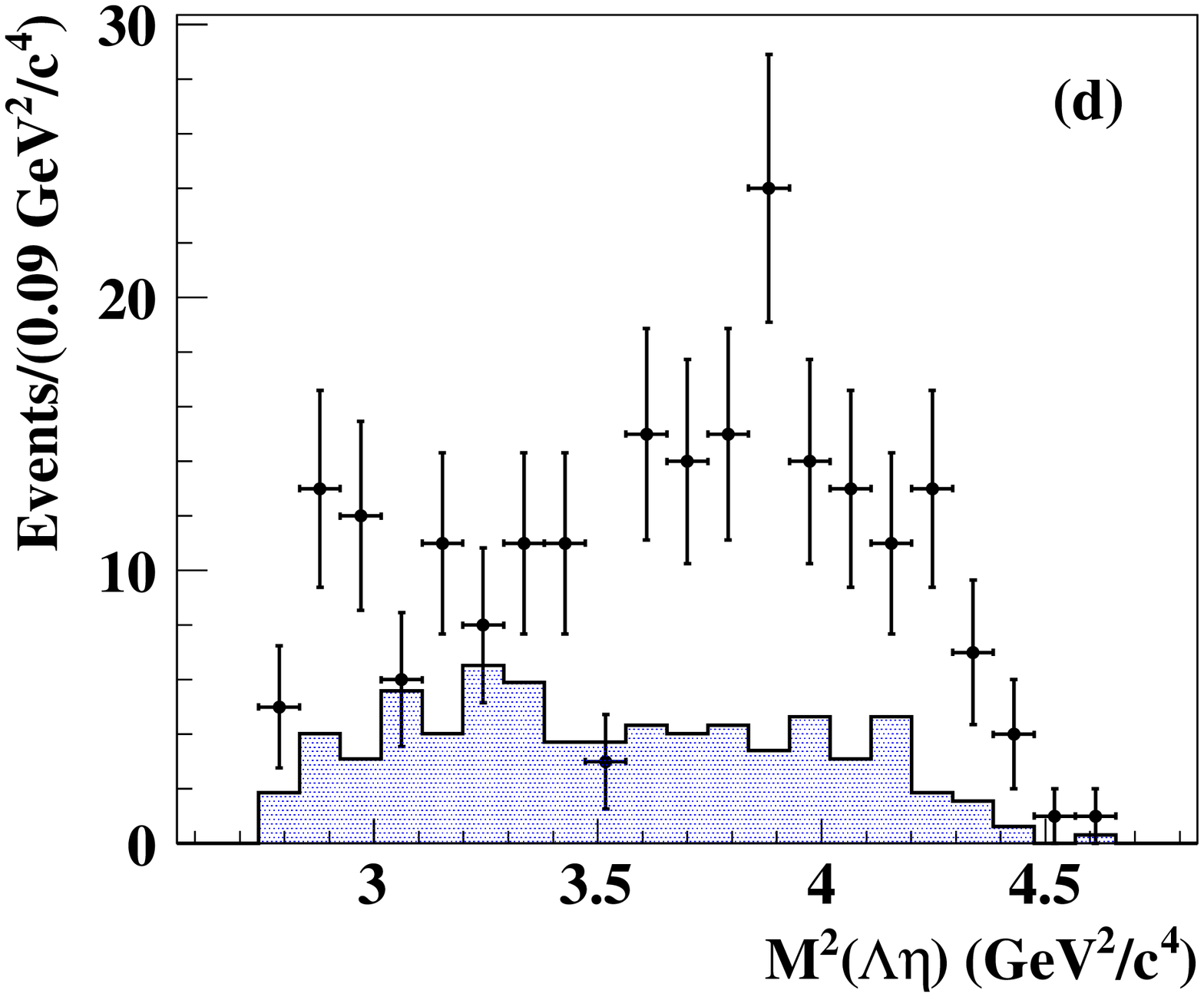}
    \includegraphics[trim = 9mm 0mm 0.mm 0mm, width=0.32\textwidth]{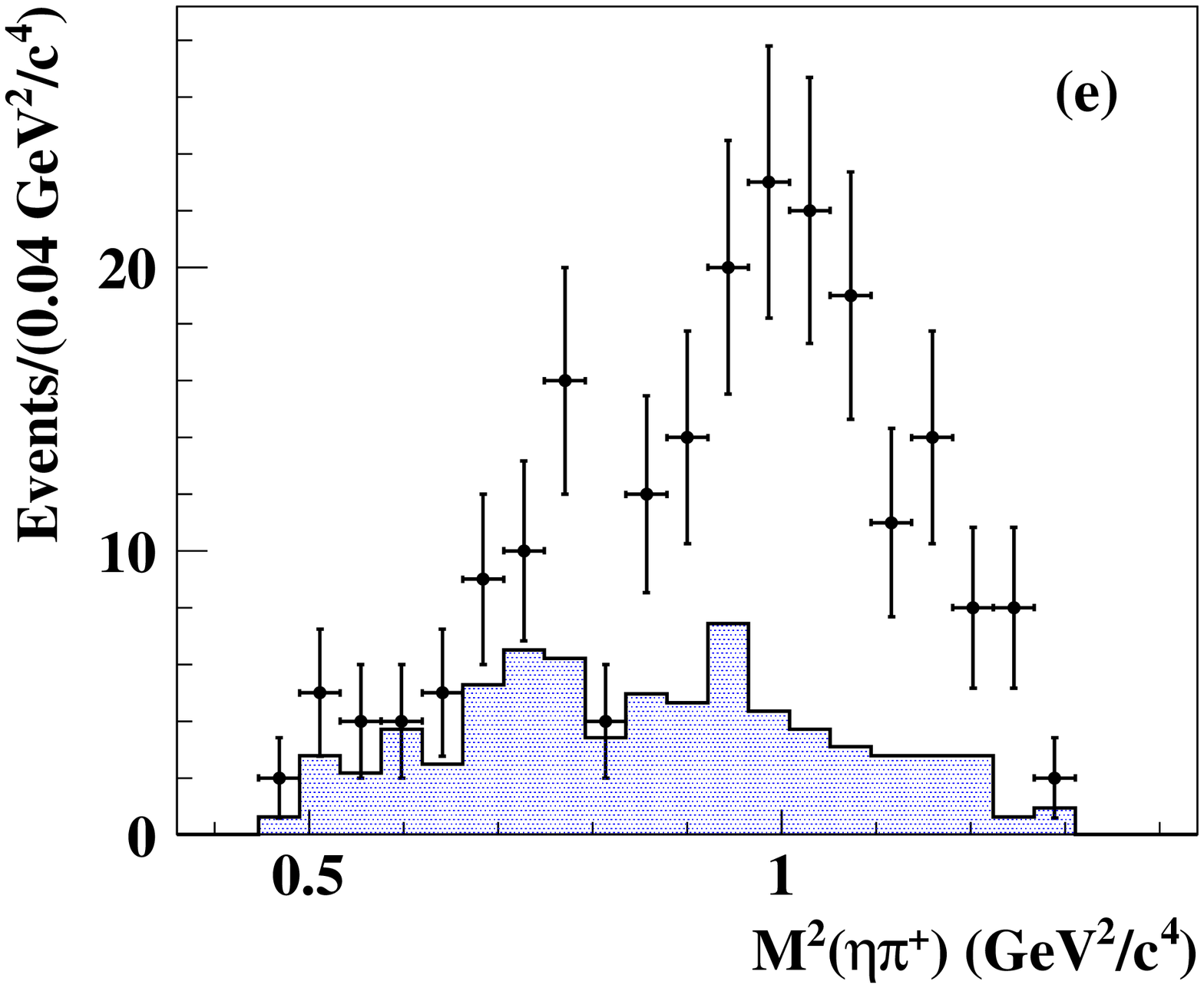}
\end{center}
\caption{Two-dimensional Dalitz distribution of $M^2(\Lambda\eta)$ versus $M^2(\Lambda\pip)$ for selected $\lmdcp\to\lmdetapi$ candidates in $\mbc$ signal (a) and sideband (not scaled) (b) regions. Also plots (c)-(e) show their one-dimensional projections, where dots with error bars stand for data in plot (a) and the shaded histograms (luminosity scaled) stand for data in plot (b).}\label{fig:dalitz}
\end{figure*}

To check the possible intermediate states fore-mentioned in the theoretical calculations~\cite{2016JJXie,2015Kenta,2017JJXie}, 
the two-dimensional Dalitz distributions of $M^2(\Lambda\eta)$ versus $M^2(\Lambda\pip)$   for selected $\lmdcp\to\lmdetapi$ candidates in the $\mbc$ signal region $2.282<\mbc<2.291\;\gevcc$ and the sideband region $2.250<\mbc<2.270\;\gevcc$ are shown in Fig.~\ref{fig:dalitz}(a) and (b), respectively.
In addition, the corresponding one-dimensional projections are presented in Fig.~\ref{fig:dalitz}(c)-(e). In the $M(\Lambda\pip)$ spectrum, an obvious peak of the $\sigstar$ resonance is seen, which has been studied at CLEO~\cite{1995RAmmar}, while other potential states are not evident in these projections. 
Hence, under the current statistics, we only measure the decay rate of $\lmdcp\to\sigstar\eta$.

\begin{figure}[htbp]
\begin{center}
	\includegraphics[trim = 10mm 0mm 0mm 0mm,width=0.4\textwidth]{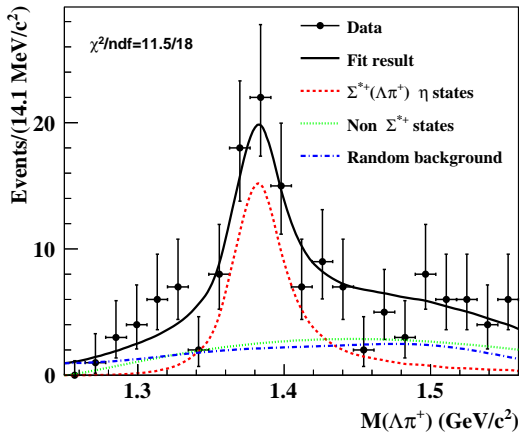}
\end{center}
\caption{Fit to the $\Lambda\pip$ invariant mass spectrum of $\lmdcp\to\lmdetapi$ candidates. The dots with error bars are the data, the (black) solid curve is fit function, which is the sum of the signal shape (red dashed curve), a smooth background shape describing the background from non $\Sigma^{*+}$ states (green dotted curve) and the shape of random combinatorial background estimated using the $\mbc$ sideband (blue dash-dotted curve). A test of goodness-of-fit with $\chi^2$ divided by the degrees of freedom is shown.}
\label{fig:1385_yield}
\end{figure}

To extract the signal yield of the cascade decay $\lmdcp\to\sigstareta$, $\sigstar\to\Lambda \pip$, an unbinned extended maximum likelihood fit is performed to the invariant mass spectrum of $M(\Lambda\pip)$ for the events within the $\mbc$ signal region.  The fitting range is $1.25<M(\Lambda\pip)<1.56\;\gevcc$ as illustrated in Fig.~\ref{fig:1385_yield}. In the fit, the signal shape is derived from the kernel-estimated non-parametric shape~\cite{keyspdf} based on signal MC samples convolved with a Gaussian function. In the Gaussian function, their parameters are allowed to vary in the fit. The signal lineshape of the $\sigstar$ is generated according the following formula 
\begin{eqnarray}
|A(m)|^2\propto \frac{q^{2L_b+1}f^2_{L_b}(q)\cdot p^{2L_d+1}f^2_{L_d}(p)}{(m^2-m_0^2)^2+m_0^2\Gamma^2(m)},
\label{eq:BW1}
\end{eqnarray}
using the mass-dependent width $\Gamma(m)$ with the expression
\begin{eqnarray}
\Gamma(m)=\Gamma_0\left( \frac{p}{p_0} \right)^{2L_d+1}\left( \frac{m_0}{m} \right)\frac{f^2_{L_d}(p)}{f^2_{L_d}(p_0)} ,
\label{eq:BW2}
\end{eqnarray}
where $m=M(\Lambda\pip)$, $m_0$ and $\Gamma_0$ are the $\sigstar$ nominal mass and width, respectively, $q$ and $p$ ($p_0$) are the daughter momenta of $\lmdcp$ and $\sigstar$ (when $\sigstar$ is at its nominal mass $m_0$) at their rest frame, respectively, and $L_b=1$($L_d=1$) is angular momentum between the two-body decay products in the $\lmdcp$($\sigstar$) rest frame. $f(p)$ are Blatt-Weisskopf barrier factors which have been detailed in Ref.~\cite{BW}. Possible interference between $\sigstar$ and non-$\sigstar$ amplitudes is neglected.
The random combinatorial background is also modeled with kernel-estimated non-parametric shape~\cite{keyspdf} based on data in the $\mbc$ sideband region. The non-$\Sigma^{*+}$ background is described with a smooth background function $f_{\rm bkg}\left(M\left(\Lambda\pip\right)\right)\propto(M\left(\Lambda\pip\right)-1.25)^c\cdot(1.75-M\left(\Lambda\pip\right))^d$,
where the parameters $c$ and $d$ are obtained from MC-simulated non-$\Sigma^{*+}$ backgrounds and fixed in the fit.
Only the integral of the signal shape in the signal region $1.32<M(\Lambda\pip)<1.45\;\gevcc$ is counted as signal yield. 
The signal yield and the corresponding detection efficiency are listed in Table~\ref{tab:sum}. The corresponding BF is calculated using Eq.~\eqref{eq:BF}, where $\eff$ is the corresponding detection efficiency and $\BR_{\rm{inter}}=\BR(\Sigma^{*+}\to\Lambda\pip) \cdot \BR(\Lambda\to p\pim) \cdot \BR(\eta\to\gamma\gamma)$ taken from the PDG~\cite{2016PDG}. The resultant BF and the corresponding statistical uncertainty are also listed in Table~\ref{tab:sum}. 

\section{\boldmath Systematic uncertainty}
	Different sources of systematic uncertainties are considered in the BF measurement, including charged particle tracking, PID, reconstruction of intermediate states, the $\deltaE$ requirement, the fitting range, the background description, the signal MC model, peaking backgrounds and intermediate BFs. 
	
	\emph{Tracking and PID for $\pip$ particle.} By studying a set of control samples of $\ee\to \pip\pim\pip\pim$ events based on data collected at energies above $\sqrt{s}=4.0\;\gev$, which are the same as used in Ref.~\cite{2015MAblikimLambdac}, the tracking and PID efficiencies are estimated in data and MC simulations. After weighting these efficiencies to the $\pip$ kinematics in the signal samples, the uncertainties associated with $\pip$ tracking and PID efficiencies are derived out to be 1.0\% for each decay mode.
	
	\emph{Reconstruction for $\Lambda$ particle.} The efficiencies for $\Lambda$ reconstruction in data and MC simulations are measured with control samples of $J/\psi\to \bar{p}K^+\Lambda$ and $J/\psi\to\Lambda\bar{\Lambda}$ events, which are the same  as studied by Ref.~\cite{RecLmd}. The uncertainties of $\Lambda$ reconstruction efficencies are estimated to be 3.7\% for each decay mode, according to the $\Lambda$ momentum and angular distributions in the signal samples.
	
	\emph{Reconstruction for $\eta$ particle.} We use a control sample of $\piz$ from $D$ meson decays~\cite{RecEta} to evaluate the $\eta$ reconstruction efficiency in the decay to two photons, taking advantage of their close kinematic phase space in the laboratory frame. By studying the control sample, the $\gamma\gamma$ reconstruction efficiencies are obtained in data and  MC simulations, and an uncertainty of 3.4\% is assigned by weighting these efficiencies to the $\eta$ momentum distribution in the signal samples.
	
	\emph{Requirement for $\deltaE$.} To estimate the systematic uncertainty arising from $\deltaE$ requirement, we repeat the measurement procedure by varying the boundaries of the $\deltaE$ signal ranges with $\pm\,1\;\mev$. The largest changes in the resultant BFs, 2.3\% and 1.5\% for the decays $\lmdcp\to\lmdetapi$ and $\lmdcp\to\sigstareta$, respectively, are taken as systematic uncertainties.
	
	\emph{Fitting range.} To estimate the systematic uncertainty associated with the fitting range, we repeat the measurements by using alternative $\mbc$ fitting ranges of $2.26<\mbc<2.30\;\gevcc$ for the decay $\lmdcp\to\lmdetapi$ and of  $1.25<M(\Lambda\pi^+)<1.55\;\gevcc$ for the decay $\lmdcp\to\sigstareta$. The changes in resultant BFs, 0.9\% and 2.7\% for the decays $\lmdcp\to\lmdetapi$ and $\lmdcp\to\sigstareta$, respectively, are considered as the systematic uncertainties.
	
	\emph{Background description.} 
	For the $\lmdcp\to\lmdetapi$ decay, we repeat the measurement by varying the $\mbc$ end-point ($2.3\;\gevcc$) in the $\uchyph=0$ARGUS function by $\pm \,0.5\;\mevcc$, by adding a Gaussian function to model the affection rising from the possible peak around $2.26\;\gevcc$ and also by using an alternative background model of a linear combination of the $\uchyph=0$ARGUS function and the MC-simulated background shape. Quadratically summing the changes in resultant BFs for these three sources brings a systematic variation of 1.8\% for $\lmdcp\to\lmdetapi$ decay. For $\lmdcp\to\sigstareta$ decay, we let the parameters of non-$\Sigma^{*+}$ background function be float and repeat the measurement procedures, which leads to a systematic change of 4.8\% on the BF result.

	\emph{Signal MC model.} For the $\lmdcp\to\lmdetapi$ decay, we consider the difference of angular and momentum distributions of final states $\Lambda$, $\eta$ and $\pip$ particles between data and signal MC samples and calculate weight factors using
	$w^i=\frac{n^i_{\mathrm {Data}}}{n^i_{\mathrm {MC}}}$,
where $i$ is a specific kinematic interval and $n$ is the number of events that pass the event selections in data or signal MC samples. The change of the re-weighted efficiency from the nominal efficiency is calculated to be 2.9\%, which is assigned as the systematic uncertainty. 
For the $\lmdcp\to\sigstareta$ decay, we calculate the polar angle $\theta_{\Sigma^{*+}}$ of the momentum of the $\Sigma^{*+}$ with respect to that of the $\lambdacp$ in the rest frame of the $\lambdacp$. We model the signal process according to the distribution of $1+\alpha\cdot\cos^2\theta_{\Sigma^{*+}}$ in the range of $-1\le\alpha\le 1$. The maximum change on the MC-determined efficiency is 1.3\%. Furthermore, we vary the nominal mass and width of the $\sigstar$ within uncertainties in PDG~\cite{2016PDG}, and the maximum change on the signal yield is 0.5\%. By summing up all contributions in quadrature, an uncertainty of 1.4\% assigned.

	\emph{Peaking background.} We estimate the sizes of the potentially underestimated peaking backgrounds by detailed background analysis of the inclusive MC samples in measurement of the  $\lmdcp\to\lmdetapi$ decay rate, which is estimated to be 1.9\%. For the studies of the $\lmdcp\to\sigstareta$ decay rate, we incorporate  complex components from non-$\Sigma^{*+}$ intermediate processes in the MC simulations of the $\lmdcp\to\lmdetapi$ decays, and analyze the amplitude of the peaking background contribution beneath the $\Sigma^{*+}$ peak. The relative peaking background rate is evaluated to be 1.6\%.
	
	\emph{Total $\lmdcp\lmdcm$ number and intermediate BFs.} In Ref.~\cite{2015MAblikimLambdac}, absolute BFs of the twelve $\lmdcp$ decay modes were measured and the total number of $\lmdcp\lmdcm$ pairs was calculated using the absolute BFs and corresponding single-tag yields. The total number is $N_{\lambdacp\lambdacm}=(105.9\pm4.8(\rm{stat.})\pm0.5(\rm{syst.}))\times10^{3}$ and corresponding uncertainty is calculated to be 4.6\% for each decay mode by adding both the statistical and systematic uncertainties in quadrature.
	The uncertainties of the intermediate BFs quoted from the PDG~\cite{2016PDG} are $\BR(\Sigma^{*+}\to\Lambda\pip)=(87.0\pm 1.5)\%$, $\BR(\Lambda\to p\pim)=(63.9\pm 0.5)\%$ and $\BR(\eta\to\gamma\gamma)=(39.41\pm 0.20)\%$, and corresponding uncertainties are calculated to be 0.9\% and 1.9\% for $\lmdcp\to\lmdetapi$ and $\lmdcp\to\sigstareta$, respectively.
	
	All these systematic uncertainties are summarized in Table~\ref{tab:uncertainties}, and the total systematic uncertainties are evaluated to be 8.4\% and 9.5\% for the $\lmdcp\to\lmdetapi$ and $\lmdcp\to\sigstareta$ decays, respectively, by summing up all contributions in quadrature.
	
  \begin{table}[htbp]
  \begin{center}
  \footnotesize
  \caption{Summary of the relative systematic uncertainties in percentage. The total values are calculated by summing up all contributions in quadrature.}
  \begin{tabular}{l c c c}
      \hline \hline
      Source         &~~~~$\lmdetapi$~~~~~& $\sigstareta$ & ~\\ \hline
      Tracking              &         1.0   & 1.0\\
      PID                   &       1.0     & 1.0\\
      $\Lambda_{p\pim}$ reconstruction  &   3.7 & 3.7\\
      $\eta_{\gamma\gamma}$ reconstruction  &  3.4 & 3.4\\
      $\Delta E$ requirement          &      2.3   &  1.5\\
      Fitting range               &  0.9 & 2.7\\
      Background description &  1.8 & 4.8\\
      Signal MC model                 & 2.9 & 1.4\\
      Peaking background & 1.9 & 1.6\\
      $N_{\Lambda_c^+\bar{\Lambda}_c^-}$    &  4.6 & 4.6\\
      $\mathcal{B}_{\rm{inter}}$             &  0.9 & 1.9\\ \hline
      Total                           &  8.4 & 9.5\\
      \hline\hline
  \end{tabular}
  \label{tab:uncertainties}
  \end{center}
  \end{table}

\section{\boldmath Summary}
In summary, the absolute branching fractions of the two processes $\lmdcp\to\lmdetapi$ and $\sigstareta$ are measured using a single-tag method on a data sample produced in $\ee$ collisions at $\sqrt{s}=4.6\;\gev$ collected with the $\uchyph=0$BESIII detector. The results are $\BR(\lmdcp \to \lmdetapi)=(1.84\pm0.21\pm0.15)\%$ and $\BR(\lmdcp \to \sigstareta)=(0.91\pm0.18\pm0.09)\%$, where the first uncertainties are statistical and the second systematic. These are the first absolute measurements of the branching fractions for these two modes, and are consistent with the previous relative measurements~\cite{1995RAmmar,2003DCronin}, but with improved precisions. 
Under the current statistics, no other potential intermediate states are concluded. 
Future $\lmdcp$ data samples with larger statistics will allow for detailed studies of the intermediate states proposed in Refs.~\cite{2016JJXie,2015Kenta,2017JJXie}.

\acknowledgments
The BESIII collaboration thanks the staff of BEPCII and the IHEP computing center for their strong support. This work is supported in part by National Key Basic Research Program of China under Contract No. 2015CB856700; National Natural Science Foundation of China (NSFC) under Contracts Nos. 11335008, 11425524, 11625523, 11635010, 11675275, 11735014; the Chinese Academy of Sciences (CAS) Large-Scale Scientific Facility Program; the CAS Center for Excellence in Particle Physics (CCEPP); Joint Large-Scale Scientific Facility Funds of the NSFC and CAS under Contracts Nos. U1532257, U1532258, U1732263; CAS Key Research Program of Frontier Sciences under Contracts Nos. QYZDJ-SSW-SLH003, QYZDJ-SSW-SLH040; the Recruitment Program of Global Experts in China; 100 Talents Program of CAS; INPAC and Shanghai Key Laboratory for Particle Physics and Cosmology; German Research Foundation DFG under Contracts Nos. Collaborative Research Center CRC 1044, FOR 2359; Istituto Nazionale di Fisica Nucleare, Italy; Koninklijke Nederlandse Akademie van Wetenschappen (KNAW) under Contract No. 530-4CDP03; Ministry of Development of Turkey under Contract No. DPT2006K-120470; National Science and Technology fund; The Swedish Research Council; U. S. Department of Energy under Contracts Nos. DE-FG02-05ER41374, DE-SC-0010118, DE-SC-0010504, DE-SC-0012069; University of Groningen (RuG) and the Helmholtzzentrum fuer Schwerionenforschung GmbH (GSI), Darmstadt; the undergraduate research program of Sun Yat-sen University.



\end{document}